# Study of Forbush effects by means of muon hodoscopes

N. S. Barbashina, V. V. Borog, A. N. Dmitrieva, R. P. Kokoulin, K. G. Kompaniets,
G. Mannocchi, A. A. Petrukhin, D. A. Room, O. Saavedra, V. V. Shutenko, D. A. Timashkov,
G. Trinchero, I. I. Yashin

*Abstract* – **Muon rate variations during Forbush effects registered by means of muon detectors DECOR, TEMP and URAGAN operated in the experimental complex NEVOD (MEPhI, Moscow) have been studied, and comparative analysis with neutron monitor data has been performed. The ratio of values of Forbush decreases in muon and in neutron fluxes is about one third, and preliminary rigidity dependence of Forbush decrease amplitude using muon data has been also obtained (for 2.4 GV cut-off rigidity). The detection of muon flux in the hodoscopic mode allows to study the dynamics of muon flux anisotropy related with magnetic field perturbations. Results of analysis of data from the new unique muon detector URAGAN indicate the change of muon flux asymmetry direction during the Forbush decrease. This phenomenon is related with the motion of solar plasma cloud and hodoscope acceptance cone relative to each other.**

## I. Introduction

A sharp decreasing of cosmic ray intensity during magnetic field perturbations in the near Earth space (Forbush effect) is one of the interesting phenomena in solar-terrestrial physics. For study of these phenomena, the world-wide net of neutron monitors located at various points of the Earth is used. Unfortunately, neutron monitors cannot give information about the directions of changes of cosmic ray flux. From this point of view, investigations of muon flux, which constitutes about 70% of charged cosmic ray particles at sea level, are very interesting and useful. If for this purpose muon hodoscopes (detectors, which can register muons from various directions simultaneously with a good angular accuracy) are used, it is possible to measure the changes in muon flux at different zenith and azimuth angles and to study directional dynamics of these changes.

In this paper the results of muon flux variation studies during the strong Forbush decreases are presented. These data were obtained by means of three muon detectors: two muon hodoscopes TEMP [1] and URAGAN [2], and coordinate detector DECOR [3].

## II. Apparatus and experimental data

Experimental complex NEVOD [4] includes three coordinate detectors (muon hodoscopes) – TEMP, DECOR and URAGAN (Fig.1).

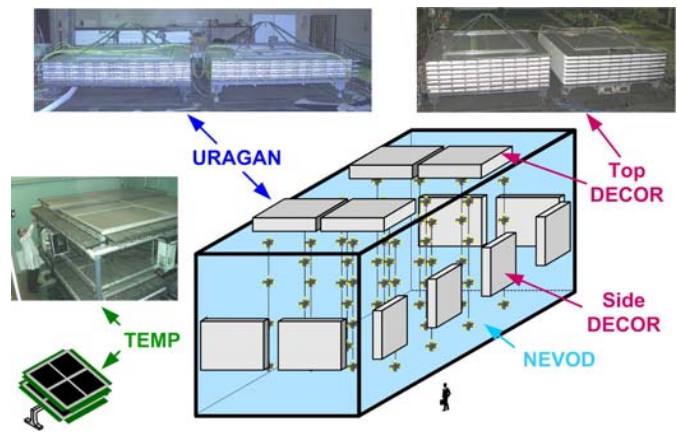

*Fig.1. Experimental complex NEVOD.*

The first in the world muon hodoscope TEMP [1] operates since 1996. TEMP consists of two pairs of horizontal coordinate planes (X, Y) with sensitive area of 9 m$^2$. These pairs are vertically separated by 1 m. Each plane is assembled of narrow scintillator counters (2.5 cm × 1 cm × 300 cm) with PMT. Total number of counters is 512; angular resolution 1 – 2°. This is nearly equivalent to recording muons with two 128 × 128 square arrays of 2.5 cm × 2.5 cm detectors. Data are continuously registered as intensity arrays with dimension 255 × 255 directional cells (the second muon hodoscope launched in Japan in 1998 [5] has 44 × 44 cells). TEMP is located in the basement of the building. Threshold muon energy is about 0.5 GeV.

In 1998–2001 the coordinate detector DECOR was deployed around the Cherenkov calorimeter NEVOD. The side part of DECOR includes eight 8-layer supermodules (SM), 8.4 m$^2$ area each, with vertical planes of streamer tube

This work was supported in part by Federal Agency for Science and Innovations, Federal Agency for Education, Government of Moscow (Department of Science and Industrial Policy, Moscow Committee of Science and Technologies), and RFBR grants 06-02-17213-a and 06-02-08218-ofi.

N. S. Barbashina, V. V. Borog, A. N. Dmitrieva, R. P. Kokoulin, K. G. Kompaniets, A. A. Petrukhin, D. A. Room, V. V. Shutenko, D. A. Timashkov, I. I. Yashin are with Moscow Engineering Physics Institute, Russia; e-mail: NSBarbashina@mephi.ru.

G. Mannocchi, G. Trinchero are with Istituto Nazionale di Astrofisica, Sezione di Torino, Italy.

O. Saavedra is with Dipartimento di Fisica Generale dell Universita di Torino, Italy.

chambers. Top supermodules of DECOR are located on the cover of the calorimeter water pool and are assembled of eight horizontal streamer tube chamber layers interlaid with 10 cm foam plastic. For the present analysis, coincidences between signals from any side DECOR supermodule and any top DECOR SM (trigger #9) are used. Such condition provides registration of muons with energy $E > 2$ GeV.

In 2005, on the basis of top DECOR supermodules a new multipurpose muon hodoscope URAGAN has been constructed. The URAGAN supermodule includes eight planes interlaid with 5 cm foam plastic and composed of 320 streamer tubes (1 cm × 1 cm × 350 cm) with external strips (along and across streamer tubes) forming two-dimensional readout system (4864 data channels). Total area of each SM is about 11.5 m$^2$. The data processing system allows to reconstruct muon tracks in the on-line mode and to register muon flux from upper hemisphere as continuous 2D-pictures. The setup provides detection of particles in a wide range of zenith angles (from 0 to 80°) with angular accuracy about 0.7°. Threshold muon energy is about 0.2 GeV.

As an example of the Forbush decrease detection with these setups, in Fig.2 variations of the counting rate during 10 - 25 May, 2005 from muon detectors and Moscow neutron monitor [6] are presented.

### III. DATA ANALYSIS

For the analysis of Forbush decreases (FD), the normalized 10-minute counting rate of three muon detectors of NEVOD complex and Moscow neutron monitor (IZMIRAN, Troitsk) were used (for the same cut-off rigidity 2.4 GV). Corrections for the barometric effect were evaluated and introduced. During the simultaneous operation of neutron monitor and muon hodoscopes (January − June 2004 and December 2004 − May 2005) nine Forbush decreases (FD) were detected; five of them with a decrease in the rate of neutrons more than 5% were selected (see Table I).

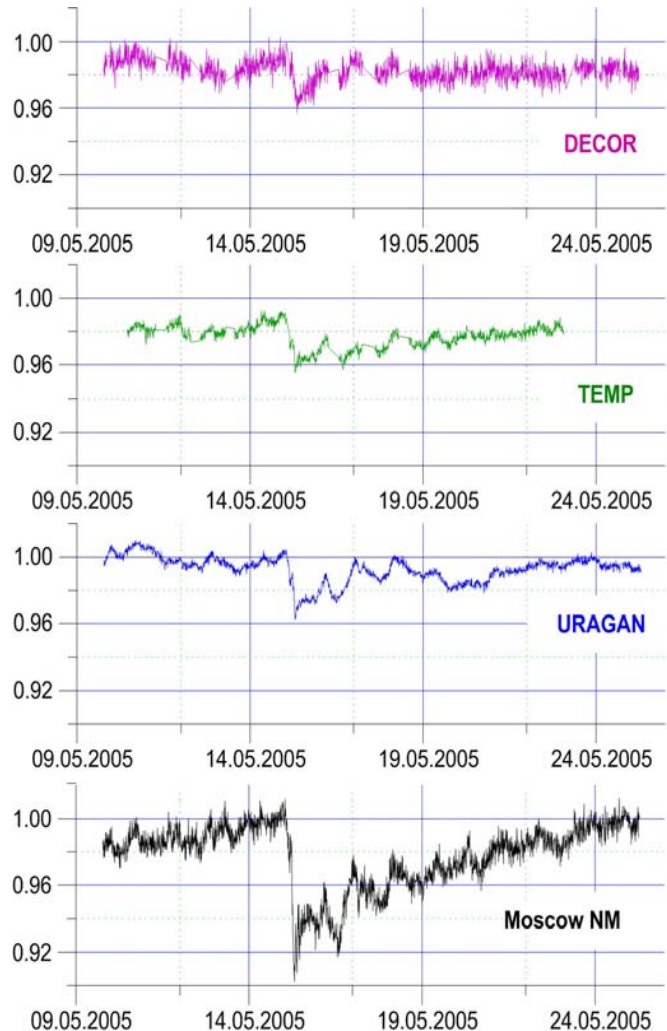

*Fig.2. Relative variations of the counting rate in May, 2005 from muon detectors DECOR, TEMP, URAGAN and Moscow neutron monitor.*

TABLE I
DECREASE IN THE RATE OF MUONS AND NEUTRONS

| FD | Moscow NM, % | URAGAN, % | TEMP, % | DECOR, % |
|---|---|---|---|---|
| 22 Jan 2004 | 7.2 | - | 2.4 | 1.3 |
| 17 Jan 2005 | 13.1 | - | 5.4 | 3.7 |
| 21 Jan 2005 | 7.3 | - | 3.0 | 1.7 |
| 8 May 2005 | 5.2 | 2.4 | - | 1.4 |
| 15 May 2005 | 7.3 | 3.3 | 2.8 | 2.2 |

The ratio of Forbush decreases in the flux of muons and neutrons is equal approximately to 1/3, since for generation of detected muons and neutrons primary cosmic ray particles of different energies are responsible.

To estimate the effective energies and rigidities of primary protons the thresholds of the setups and muon energy loss in the atmosphere were taken into account. The coupling muon-proton energy coefficient was taken as 10. The dependence of FD amplitude on thus calculated proton rigidity for five events listed in Table I is shown in Fig.3. Amplitudes of Forbush decrease are well fitted by power dependence with the exponent identical for all five Forbush effects:

$$A_{FD} \sim R_p^{-0.9 \pm 0.15}.$$

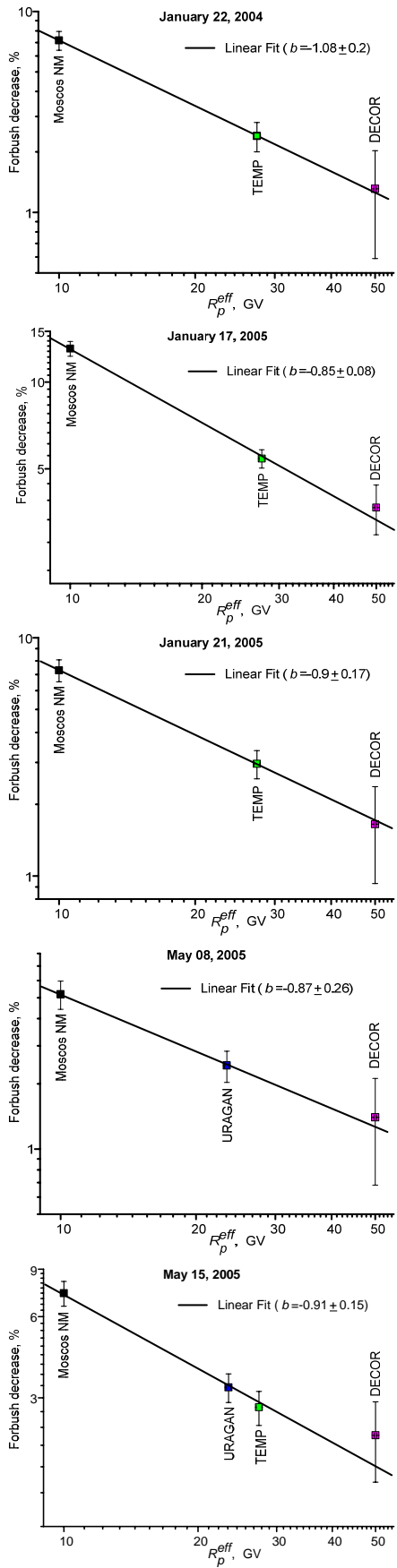

*Fig.3. Dependence of FD amplitude on effective rigidity of primary protons.*

The new muon detector URAGAN launched in March 2005 allows to investigate a spatial-angular structure of muon flux. In Fig.4, the preliminary URAGAN data on anisotropy of muon intensity during the Forbush decrease of May 15, 2005 are shown. In the bottom panel relative deviation of muon flux from the average one as a function of azimuth angle φ is plotted. The results have been obtained by summation of muon numbers from 25° to 65° zenith angles Coordinate system is attached to the NEVOD calorimeter with X-axis directed to South-West. From the figure, a steady gradual shift of muon flux asymmetry direction approximately by 60 degrees (from 180° to 120°) is seen. Such variations can be explained by the Earth rotation (during three hours) relative to solar plasma cloud coming to the Earth.

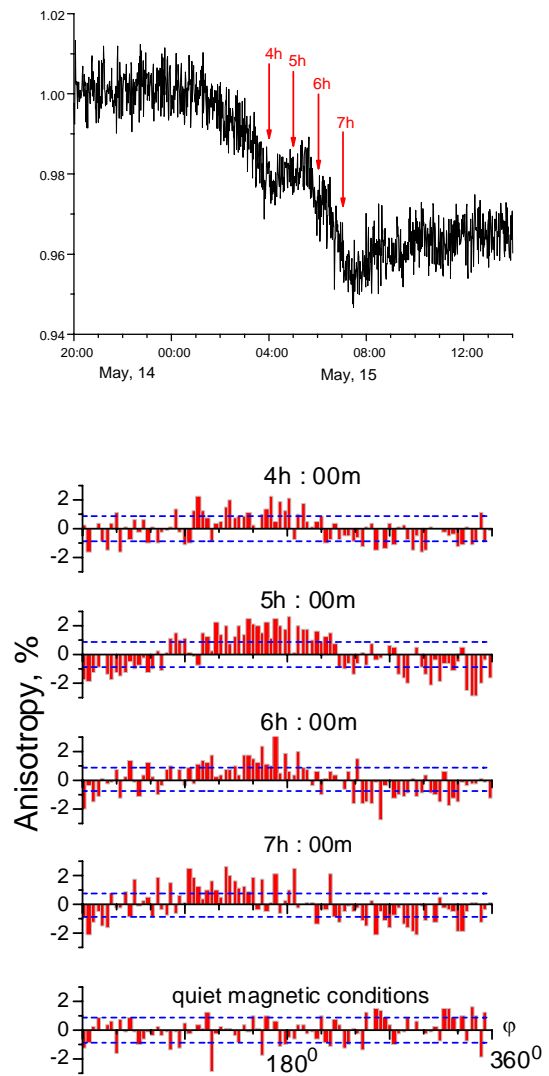

*Fig. 4. Preliminary URAGAN data during Forbush decrease of May 15, 2005: total muon rate (top panel), anisotropies of muon intensity (bottom panel).*

## IV. Conclusion

The system of three independent muon detectors operating in one experimental complex gives a unique possibility to investigate muon flux variations during FD in high energy primary particles region and to study various processes in heliosphere and the Earth's magnetosphere using rigidity dependencies, angular anisotropies and others properties of cosmic ray flux dynamics.